\documentclass[submitted]{IEEEtran}
\usepackage{graphicx}
\usepackage{url}
\usepackage{cite}
\usepackage{color}
\usepackage{upgreek}
\usepackage[colorlinks=true,linkcolor=blue,urlcolor=blue,citecolor=blue]{hyperref}
\usepackage{changes}
\begin{document}

\title{\bf Updates on the Tsinghua Tabletop Kibble Balance}
\author{Shisong Li, Yongchao Ma, Kang Ma, Weibo Liu, Nanjia Li, Xiaohu Liu, \\ Lisha Peng, Wei Zhao, Songling Huang, Xinjie Yu

\thanks{Shisong Li, Yongchao Ma, Kang Ma, Weibo Liu, Nanjia Li, Xiaohu Liu, Lisha Peng, Wei Zhao, Songling Huang, and Xinjie Yu are with the Department of Electrical Engineering, Tsinghua University, Beijing 100084, China. Wei Zhao is also with the Yangtze Delta Region Institute of Tsinghua University, Jiaxing, Zhejiang 314006, China.}
\thanks{This work was supported by the National Natural Science Foundation of China under Grant 52377011 and in part by the National Key Research and Development Program of China under Grant 2022YFF0708600.}
\thanks{Email: shisongli@tsinghua.edu.cn}}

\maketitle

\begin{abstract}
With the adoption of the revised International System of Units (SI), the Kibble balance has become a pivotal instrument for mass calibrations against the Planck constant, $h$. One of the major focuses in the Kibble balance community is prioritizing experiments that achieve both high accuracy and compactness. The Tsinghua tabletop Kibble balance experiment seeks to develop a compact, high-precision, user-friendly, cost-effective, and open-hardware apparatus for mass realization, specifically within the kilogram range. This paper reports on the progress of the Tsinghua tabletop Kibble balance project over the past two years. Various aspects of the Tsinghua tabletop system, including electrical, magnetic, mechanical, and optical components, are summarized. Key achievements, such as the construction and characterization of the magnet system, determination of absolute gravitational acceleration, investigation of a capacitor-sensor-based weighing unit, and development of a high-precision current source, are presented to provide a comprehensive understanding of the experiment's status.
\end{abstract}


\section{Introduction}

\IEEEPARstart{S}{ince} May 20, 2019, fundamental metrology has ushered in a quantum era with the adoption of the new International System of Units, SI~\cite{cgpm2018}. Within the field of mass metrology, the Kibble balance, an experiment initially proposed by Dr. Br{y}an Kibble from the National Physical Laboratory (NPL, UK) in 1976~\cite{Kibble1976}, becomes one of the two primary approaches capable of calibrating a mass $m$ against the Planck constant, $h$~\cite{haddad2016bridging}. Another significant approach to mass realization involves the x-ray crystal density (XRCD) method~\cite{bartl2017new}, also known as the International Avogadro Coordination (IAC) project.

At present, a considerable number of metrology institutes, including \cite{NIST,NRC,NPL3,BIPM,LNE,METAS,NIM,KRISS,UME,MSL,PTB}, are actively engaged in conducting the Kibble balance experiment for mass realizations, primarily focusing on the kilogram level mass calibration.  A best relative measurement uncertainty of $1\times10^{-8}$ is achievable for a Kibble balance \cite{NRC,NIST}.  However, highly accurate Kibble balance systems are usually also very sizable and complex to operate.  A noteworthy development in recent times is the advent of tabletop Kibble balances that manage to uphold both precision and compactness, attracting widespread attention from the metrology community \cite{NPL3,PTB,chao2020performance,li2022design}.  It is evidenced in \cite{li2022} that tabletop Kibble balance systems may suffer from uncertainty components inversely related to the system size.  It is therefore still challenging work to explore measurement uncertainty boundaries in compact Kibble balance systems.  

Against this backdrop, Tsinghua University launched a tabletop Kibble balance project in late 2022 \cite{li2022design}. The primary objective of this initiative is to produce a compact, highly accurate, user-friendly, and cost-effective mass realization instrument, adhering to an open-hardware approach. Open hardware \cite{li2022designTIM} means openly accessible design, and experimenters are allowed to copy and build the same system with a limited number for non-commercial purposes.

This paper, as an extension of the digest presented at the 2024 Conference on Precision Electromagnetic Measurements (CPEM) in Denver, USA \cite{THU2024cpem}, aims to encapsulate the ongoing progress of the Tsinghua tabletop Kibble balance experiment, offering insights into the advancements made in various aspects of its design and implementation. The paper is organized as follows: In section \ref{sec:02}, a brief introduction to the Tsinghua system together with overall progress is presented. Then, the design and realization of some major components of the experiments, including the magnet system, weighing unit, electrical current source, gravity measurement, and optical design, are discussed in section \ref{sec:03}. Finally, a conclusion is drawn in section \ref{sec:04}.

\section{Overall setup of the experiment}
\label{sec:02}

\begin{figure}[t!]
    \centering
    \includegraphics[width=0.42\textwidth]{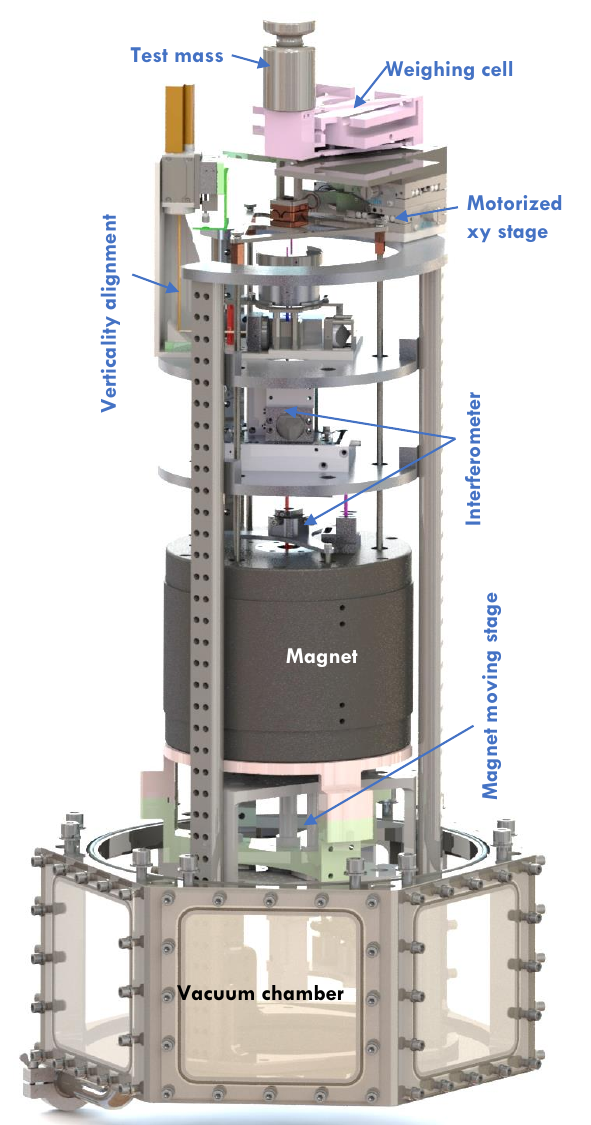}
    \caption{CAD model of the updated Tsinghua tabletop Kibble balance.}
    \label{fig:01}
\end{figure}

The measurement target of the Tsinghua tabletop Kibble balance is to calibrate masses ranging from 10\,g to 1\,kg with a measurement uncertainty of below 50\,$\upmu$g. At the present stage, a highly accurate mass realization at 1\,kg is the major focused task. Starting from the original design proposed in 2022 \cite{li2022design}, iterations of the design have been made, and the latest CAD model of the experimental apparatus is shown in Fig.~\ref{fig:01}. The total size (vacuum chamber) is about 1\,m in height and 0.35\,m in diameter. The magnet system is a core element of optimization, and the design allows the production of a high $Bl$ value with low coil-ohmic heating, meanwhile, a significant volume and mass reduction (A summary of details is presented in \ref{mag}). The magnet-moving mechanism is used in the velocity measurement. The magnet moving range is $\pm$5\,mm, and the weighing unit is fixed during the weighing phase to ensure its measurement precision. The drawback of the magnet-moving mechanism is a background flux issue~\cite{UME,NIM}, i.e., background or external magnetic flux sources can present a residual static field at the coil position due to finite self-shielding, which can not be detected by the velocity measurement but is seen by the weighing measurement. The Tsinghua tabletop Kibble balance employs a one-mode, two-phase (OMTP) measurement scheme~\cite{BIPM}. In the weighing phase, the magnet's position is fixed, and mass-on and mass-off measurements are carried out, i.e.
\begin{eqnarray}
    BlI_++\frac{I_+^2}{2}\frac{\partial L}{\partial z}&=&m \textsl{g}-m_c \textsl{g}+\Delta f_+, \nonumber\\
    BlI_-+\frac{I_-^2}{2}\frac{\partial L}{\partial z}&=&-m_c \textsl{g}+\Delta f_-,
    \label{eq:weighing}
\end{eqnarray}
where $I_+$, $I_-$ are currents through the coil with mass-on and mass-off; $Bl$ is the magnetic geometical factor; $m$ is the mass to be calibrated, and $\textsl{g}$ the local gravitational acceleration; $m_c \textsl{g}$ denotes the counter weight; $L$ is the inductance of the coil, $\partial L/\partial z$ is its gradient along the vertical $z$. $\Delta f_+$ and $\Delta f_-$ are the residual forces measured by the weighing unit. A combination and {simplifications} of (\ref{eq:weighing}) yields
\begin{equation}
    Bl=\frac{m \textsl{g}}{I_+-I_-}+\frac{\Delta f_+-\Delta f_-}{I_+-I_-}-\frac{\partial L}{\partial z}\frac{I_++I_-}{2}.
    \label{eq:Blw}
\end{equation}

Ideally, a symmetrical current setup, i.e. $I_++-I_-=0$, and a high-precision null force detector, i.e. $\Delta f_+-\Delta f_-=0$, { can ensure the elimination of the last two terms in (\ref{eq:Blw}).} In reality, how well these two terms are suppressed depends on the performance of the weighing unit and the {treatment} of the current effect~\cite{li17,li2022irony}. 
{ The OMTP scheme's velocity phase includes measurements with positive and negative currents.} The equations are written as
\begin{eqnarray}
    \left(Bl+I_+\frac{\partial L}{\partial z}\right)v_+=U_+, \nonumber\\
    \left(Bl+I_-\frac{\partial L}{\partial z}\right)v_-=U_-,
    \label{eq:velo}
\end{eqnarray}
The average of two $Bl$ profiles in (\ref{eq:velo}) yields
\begin{equation}
    Bl=\frac{\displaystyle\frac{U_+}{v_+}+\frac{U_-}{v_-}-\displaystyle\frac{\partial L}{\partial z}(I_++I_-)}{2}.
    \label{eq:Blv}
\end{equation}
A symmetrical current setup can help reduce the current dependence of $Bl$ and simplify the data processing. A combination of (\ref{eq:Blw}) and (\ref{eq:Blv}) with the current effect ($\partial L/\partial z$ terms) well compensated can yield
\begin{equation}
m=\frac{(\frac{U_+}{v_+}+\frac{U_-}{v_-})(I_+-I_-)}{2 \textsl{g}}-\frac{\Delta f_+-\Delta f_-}{\textsl{g}}.
\label{eq:kb}
\end{equation}

The equation (\ref{eq:kb}) delineates the fundamental measurement principle of the Tsinghua Kibble balance.  Illustrated in Fig.~\ref{fig:01}, the recent two years have witnessed notable progress in the Tsinghua tabletop Kibble balance, with a particular emphasis on enhancing the measurement precision of the parameters outlined in (\ref{eq:kb}).  A magnet system is built to offer a precision link between the weighing and velocity measurements.  The gravitational acceleration measurement determines $\textsl{g}$. The weighing unit investigation aims to ensure the precision of $\Delta f$. A two-stage current source is proposed to supply a precision $I$. The optical design is to ensure a good measurement accuracy of $U/v$. The development of these key components is presented in \ref{sec:03}.

\section{Progress on some key components} 
\label{sec:03}
\subsection{Magnet System}
\label{mag}
Two different magnet systems for the Tsinghua Kibble balance have been constructed, utilizing Sm$_2$Co$_{17}$ and NdFeB materials. The design, manufacture, assembly, and testing processes are detailed in \cite{li2024magnet}. The magnets are compact, with a maximum outer diameter of 220\,mm and a height of 180\,mm. Each magnet weighs approximately 40\,kg. The magnetic field strength achieved in a 15\,mm-wide air gap is 0.44\,T for Sm$_2$Co$_{17}$ and 0.59\,T for NdFeB. Since the OMTP measurement can well balance the coil ohmic heating during the weighing and velocity measurements, the NdFeB magnet will be used in the Tsinghua experimental system due to its higher $Bl$ value. The Tsinghua system employs a bifilar coil wound with 0.2\,mm gauge wire. The coil has a mean radius of 80.5\,mm, with a sectional wire region 15\,mm in height and 11\,mm in width, comprising 1360 turns. In this configuration, the $Bl$ is approximately 400\,Tm. The resistance of a single coil is about 400\,$\Omega$.

\begin{figure}
    \centering
    \includegraphics[width=0.5\textwidth]{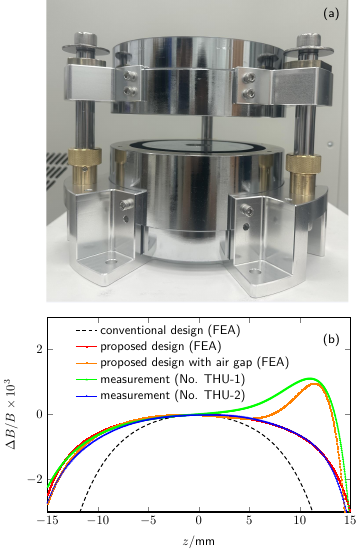}
    \caption{(a) the magnet system for the Tsinghua tabletop Kibble balance. The plot shows the open status of the magnet, where the upper segment of the magnet is levitated by the magnetic force. In real Kibble balance operations, the upper and lower segments stay in closed status. (b) {presents} the magnetic profile of the Tsinghua tabletop Kibble balance magnet. The measurement result is obtained at a mean radius of $r_\mathrm{a}=81.5$\,mm. }
    \label{fig:magnet}
\end{figure}

A novel feature of the Tsinghua Kibble balance magnet is its easy splitting operation. As shown in Fig.~\ref{fig:magnet}(a), the magnet is divided into upper and lower segments by a horizontal open/close surface located 15\,mm above the horizontal symmetrical plane ($z=0$\,mm). This design creates a unique interacting force between the two segments: when the splitting distance $d$ is within a few millimeters, the magnetic force between the segments is attractive, reaching a maximum of 200\,N at $d=0$\,mm. Combined with the {self-weight} of the upper segment (170\,N), this total force maintains the closed position.
Once the separation distance exceeds a few millimeters, the magnetic force direction reverses, and the repulsive force reaches a peak of 540\,N at $d=6$\,mm, then gradually decreases as $d$ increases further. Since the {weight} of the upper segment (170\,N) is much lower than the maximum repulsive force, the upper segment can be robustly levitated by the magnetic force when in the open position, as shown in Fig.~\ref{fig:magnet}(a). To close the magnet, a downward pushing force of over 370\,N is needed, which can be easily applied by hand.
The separation distance in the levitated state, where the gravity of the upper segment counterbalances the repulsive magnetic force, is approximately 40\,mm. This distance is ideal for coil operations, such as inserting or removing the coil and modifying optical elements. A guiding system allows for in-situ open/close operations of the magnet system, significantly reducing the maintenance costs of a Kibble balance experiment.

The second major feature of the Tsinghua magnet system is the inner yoke compensation for achieving a much wider uniform field range along the vertical $z$. The idea was originally proposed in \cite{li2020simple}, and this compensation is now for the first time applied in practice. The rectangle shape added to the two ends of the inner yoke has a {thickness} of 0.4\,mm and a height of 5\,mm. This simple modification causes almost no additional cost in manufacturing, but it can greatly improve the field uniform range. Fig.~\ref{fig:magnet}(b) shows the magnetic profiles with and without inner yoke compensations. It can be seen from Fig.~\ref{fig:magnet}(b) that the uniform field range increases by over 50\% with the compensation, verified by the experimental measurement result. Note that it is observed in the measurement that the profile may be affected by the unavoidable air gap between the upper and lower segments. The No. THU-1 result is an example with a tiny gap of 0.14\,mm and a considerable ripple appears above the splitting surface. This can be fixed by using {a thinner Nickel-coating rust protective layer of the yoke and tightening the screws locking the upper and lower segments, which can reduce the magnetic reluctance on the open/close surface and thus improve the profile uniformity, e.g. No. THU-2.}
The design, manufacture, test tools, and performance evaluation of the Tsinghua Kibble balance magnet will be released and go open-hardware once our magnet paper \cite{li2024magnet} is published. 

\subsection{Gravitational Acceleration Measurement}

\begin{figure}[tp!]
    \centering
    \includegraphics[width=0.5\textwidth]{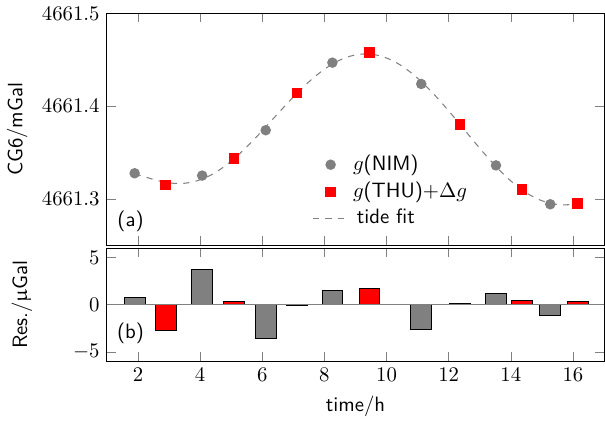}
    \caption{(a) shows the measurement result of the gravitational acceleration using the relative gravimeter, the CG6. During the measurement, the automatic tide corrections of CG6 are shut down and hence a correction or a fit should be applied to remove the time-varying effect. The fit shown in the plot is a referenceless 6-order polynomial fit proposed in \cite{THU2024gravity}. (b) presents the residual of the measurement after the tide correction is applied. The standard deviations for the NIM and THU measurements are respectively 2.5\,$\upmu$Gal and 1.5\,$\upmu$Gal. }
    \label{fig:g}
\end{figure}

The absolute gravitational acceleration at the Tsinghua tabletop Kibble balance site was determined as detailed in \cite{THU2024gravity}. This determination involved transferring values from an internationally recognized key comparison site at the Changping campus of the National Institute of Metrology (NIM) in China \cite{wu2020results}, utilizing a relative gravimeter, the CG6. The gravitational acceleration at the Tsinghua Kibble balance site (THU) and the NIM reference site (NIM) was alternately measured in seven loops over 14 hours, with the final measurement results shown in Fig.~\ref{fig:g}(a).
Two data analysis approaches were used: tide corrections with a superconducting relative gravimeter (iGrav-SG) and a referenceless polynomial fitting method. These methods yielded \textit{g} differences between THU and NIM of $(16~658.4\pm2.1)\,\upmu$Gal and $(16~657.6\pm1.2)\,\upmu$Gal, respectively. Combining the two results, the absolute gravitational acceleration at the reference point in the Tsinghua Kibble balance room was found to be \textit{g}$_0=(980~139~580.5\pm1.4)\,\upmu$Gal.
Notably, the noise level at the Kibble balance site, located on the fourth floor of the west main building at Tsinghua University, appears to be comparable to that at the NIM reference point, as illustrated in Fig.~\ref{fig:g}(b).

Additionally, gravity gradients in both the horizontal (HGG) and vertical (VGG) directions were mapped using relative gravity measurements by CG6. Given that the balance site is relatively small (1.9\,m $\times$ 2.5\,m), no significant HGG was detected. However, the VGG was precisely determined to be $\partial\textit{g}/\partial z = -296.2\,\upmu$Gal/m.
Furthermore, several additional effects were modeled, including the Earth's tide correction, the atmospheric mass effect, the polar motion effect, and the self-attraction effect. The uncertainty in the final determination of the absolute gravitational acceleration is 5.4\,$\upmu$Gal ($k=2$).

\subsection{Weighing Unit}

In order to achieve accurate weighing measurements, a high-performance force comparator is essential. The desired weighing unit must meet the following criteria: 1) a weighing resolution at the microgram level, 2) a dead load capacity of several kilograms, and 3) an active servo control capacity of several tens of grams. While commercial weighing units may meet these specifications, modifying their hardware and PID parameters to adapt to the Kibble balance method is challenging. Therefore, in the Tsinghua Kibble balance project, a custom-designed weighing cell has been developed to satisfy these specific requirements. 
\begin{figure}[tp!]
    \centering
    \includegraphics[width=0.5\textwidth]{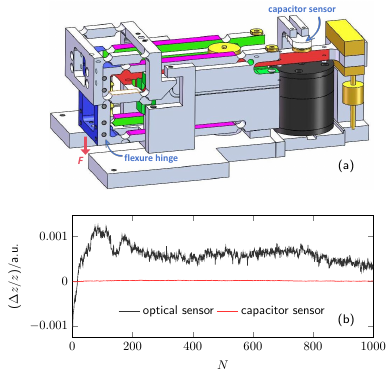}
    \caption{(a) presents the weighing unit design used in the Tsinghua tabletop Kibble balance. (b) shows a comparison of the relative resolution of a capacitor sensor and an optical sensor.}
    \label{fig:weighingcell}
\end{figure}
The design of the weighing unit is illustrated in Fig.~\ref{fig:weighingcell}(a). {With sufficient current measurement accuracy ($10^{-9}$ using a Programmable Josephson Voltage Standard (PJVS) and a standard resistor), the weighing sensitivity of the unit is determined as:  
\begin{equation}
\Delta F=\frac{\partial F}{\partial z}\Delta z,
\end{equation}
where ${\partial F}/{\partial z}$ represents the stiffness of the flexure hinge, and $\Delta z$ corresponds to the displacement measurement resolution.  
Optical sensors are commonly employed in commercial weighing cells for $\Delta z$ measurements. However, a key innovation in the design shown in Fig.~\ref{fig:weighingcell}(a) is the incorporation of a capacitive sensor for position measurement. This sensor has a measurement range of a few hundred micrometers and is used as the feedback control target for the current-carrying coil. The primary objective of this design is to enhance the sensitivity of $\Delta z$ detection, enabling the use of a stiffer flexure hinge.  
Fig.~\ref{fig:weighingcell}(b) illustrates a comparison between the output of the capacitive sensor and that of an optical position sensor when the flexure beam is held at a fixed position. Both signals are normalized relative to their respective measurement ranges, which span a few hundred micrometers. It can be seen that compared to the optical sensor, the capacitor sensor demonstrates superior stability and significantly improved position measurement resolution. This enhancement in position detection sensitivity can reduce the sensitivity requirement of the flexure hinge, allowing for the use of a thicker hinge, which in turn increases the dead load capacity.

The first prototype of the weighing unit has been constructed. The flexure hinge size is approximately 120\,mm\,$\times$\,40\,mm\,$\times$\,50\,mm, with the thinnest flexure (at the rotation center) having a thickness of 60\,$\upmu$m. Experimental measurements indicate a stiffness of ${\partial F}/{\partial z} \approx 275$\,N/m, exhibiting excellent linearity. The prototype utilizes a capacitive displacement sensor with a measurement range of 400\,$\upmu$m and a resolution of $\Delta z = 0.35$\,nm. The servo control loop is operational, and testing of the overall weighing performance is currently underway. The final goal of the weighing unit for tabletop Kibble balances is to achieve force repeatability below 10\,$\upmu$g and a total load capacity (including dead weight) of 3\,kg. Note that in Fig.~\ref{fig:weighingcell}(a), the black cylinder represents the auxiliary magnet system used for weighing performance tests. This magnet-coil system will eventually be replaced by the Kibble balance coil and magnet developed in \cite{li2024magnet}.

\subsection{Current Source}

Under the OMTP measurement scheme, the current source should offer precision current output for both the weighing and velocity measurements. The targets of the current source for the two measurement phases are different: As shown in Fig.~\ref{fig:CircuitDiagram}(a), in the weighing measurement, the current should be able to supply the required resolution for the servo control loop, and the coil position is the control target. While in the velocity measurement phase, as shown in Fig.~\ref{fig:CircuitDiagram}(b), in order to avoid the electrical noise caused by the current effect ($\partial I/\partial t$ term), the current source should remain stable. 

\begin{figure}[htbp!]
    \centering
    \includegraphics[width=0.4\textwidth]{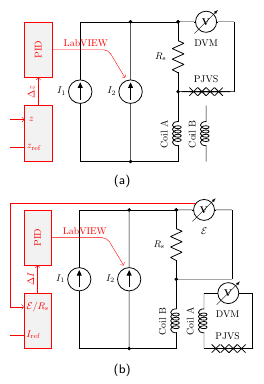}
    \caption{{The schematic diagram illustrates the two measurement phases of the Tsinghua tabletop Kibble balance: (a) weighing measurement and (b) velocity measurement. Coil A and Coil B represent the two coils of the bifilar coil, and $R_\mathrm{s}$ denotes a $100\,\Omega$ standard resistor. The system incorporates two current sources: $I_1$, a large-range current source, and $I_2$, a high-resolution, small-range current source.  In the weighing phase, the coil position $z$ is referenced to a set position $z_\mathrm{ref}$, and their deviation is fed back to $I_2$ via a LabVIEW-based PID control system. In the velocity phase, the current is switched to Coil B. The current is measured as the voltage drop across the standard resistor, $\mathcal{E}/R_\mathrm{s}$, and compared to a set value, $I_\mathrm{ref}$. The deviation from $I_\mathrm{ref}$ is servo-controlled by another PID loop to ensure the short-term stability. }}
    \label{fig:CircuitDiagram}
\end{figure}

In the Tsinghua tabletop Kibble balance, a simple bipolar current source with an output capacity of $\pm$20\,mA and stability of a few nA/A has been developed~\cite{THU2024currentsource}, and the schematic diagram is shown in Fig. \ref{fig:CircuitDiagram}. The design merges two commercially available current sources: The first stage {current source $I_1$}  is a Keithley 2410, and the range is set to 20\,mA with an output resolution 0.5\,$\upmu$A. The second stage {$I_2$} is a Keithley 6221, and the range is set to 2\,$\upmu$A with a resolution of 0.1\,nA. The idea is to use the second stage source {$I_2$} to compensate for the output fluctuation, { i.e. the current $I_1$ in the velocity phase and the magnetic force $BlI_1$ in the weighing phase.} The servo control of the proposed two-stage current source is realized by a LabVIEW program. Note that { closed} loops are used only in the velocity measurement { to achieve a highly stable current output}, and the current source in the weighing measurement works in an open loop as part of the force control loop {where the short-term stability of the output current is less important. Fig.~\ref{fig:current}(a) compares the output stability with and without the second stage source $I_2$ in the measurement scheme presented in Fig.~\ref{fig:CircuitDiagram}(b). Fig.~\ref{fig:current}(b) presents the Allen deviation of two measurement results.} It can be seen that the fluctuation of a single current source is significantly reduced and the stability of the output can reach nA/A level in about 30 minutes. 

\begin{figure}[tp!]
    \centering
    \includegraphics[width=0.5\textwidth]{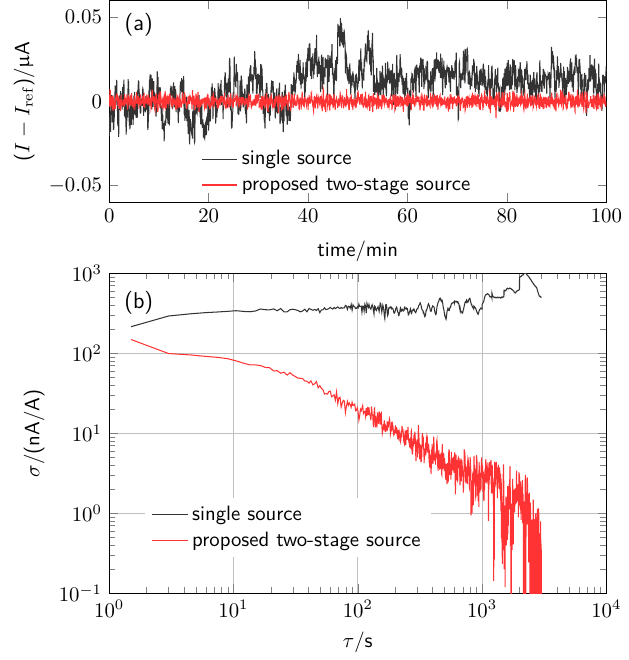}
    \caption{(a) depicts the output fluctuation of the current source with and without second-stage source compensation, where $I$ represents the current measured as $\mathcal{E}/R_\mathrm{s}$, and $I_\mathrm{ref}$ is the target value set to 12.5\,mA. (b) illustrates the Allan deviation of the measured current relative to $I_\mathrm{ref}$.} 
    \label{fig:current}
\end{figure}

\subsection{Optical System}

The magnet-moving mechanism employed in the Tsinghua tabletop Kibble balance incorporates a motorized stage currently undergoing testing, boasting a substantial load capacity of approximately 50\,kg. The relative displacement between the magnet and the coil is precisely gauged through the utilization of a frequency-stabilized laser, ZYGO7724. As shown in Fig.~\ref{fig:optical}, an interferometric measurement scheme, originally proposed by the BIPM Kibble balance group~\cite{bielsa2021new}, is employed. The coil concern cube is attached to the coil former through a hard connection of three aluminum rods. The coil cube can be adjusted along the $xy$ plane so that an optical center, where the $z$ displacement is most insensitive to the coil rotation, can be obtained. The magnet cube is mounted on the top surface of the magnet and a multiple-freedom adjustment mechanical system is used to ensure the movement of the magnet along a straight line. The interferometer measures the displacement between the coil and the magnet. Two optical signals are firstly converted into electrical outputs with photo receivers and then the electrical signals are input into a time interval analyzer (TIA). Finally, the TIA and the digital voltmeter (DVM, for measuring the induced voltage during the velocity measurement) are triggered by a master FPGA to synchronize the $U$ and $v$ measurements so that the common noise can be well canceled out \cite{haddad2015first}. {The optical system includes a mechanism for vertical alignment}, shown in Fig.~\ref{fig:optical}. The principle is similar to \cite{zeng2023vacuum} but with a more compact design. The optical systems are under test at present. 

\begin{figure}[tp!]
    \centering
    \includegraphics[width=0.42\textwidth]{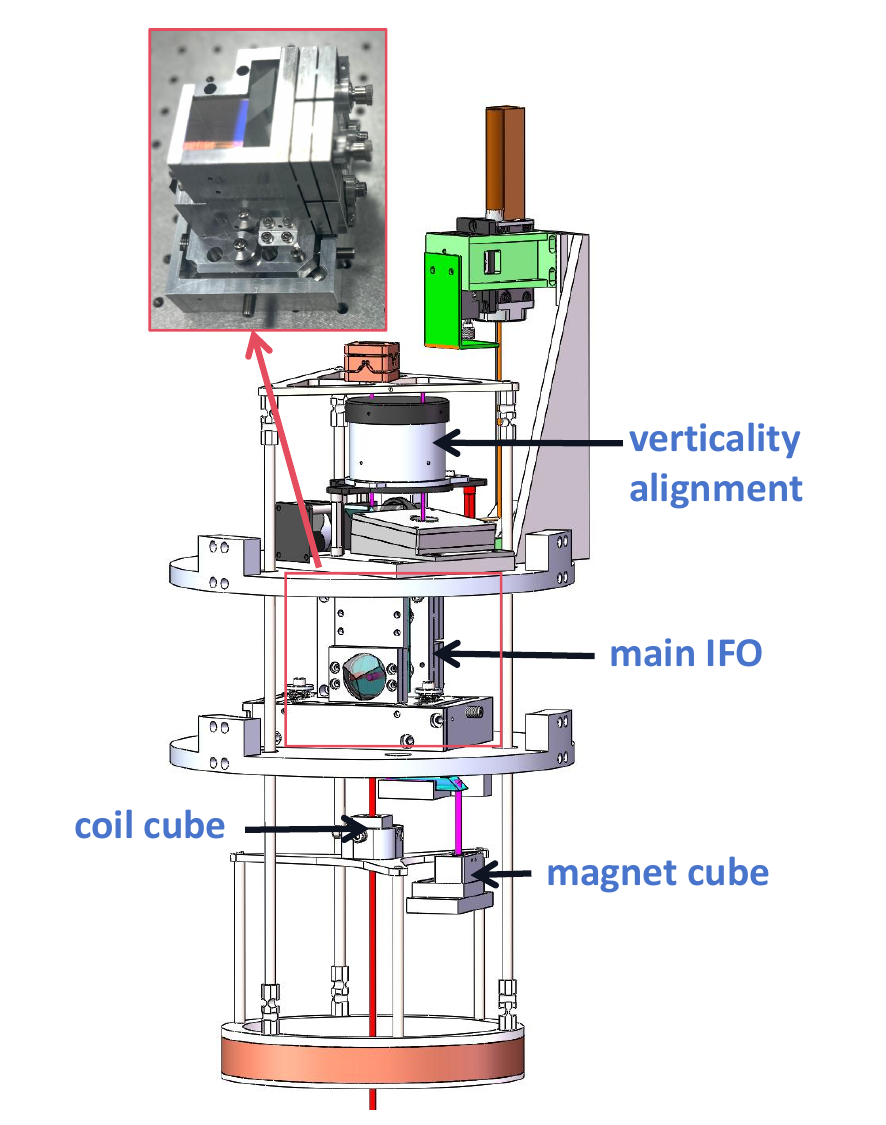}
    \caption{Design of the interferometer system for the Tsinghua tabletop Kibble balance.}
    \label{fig:optical}
\end{figure}

\subsection{Supporting standards}
For a Kibble balance, many high-precision standards and references are required to support different calibrations and measurements. The following is the progress of some related standards. 
\begin{itemize}
    
    \item [-] \textit{Voltage standard for $U$ and $V$ measurements.}
    The operation of the Kibble balance necessitates the use of a quantum voltage standard. To fulfill this requirement, we have initiated the procurement process for a 10\,V Josephson voltage system. This system serves as a crucial component, supplying reference signals integral for the accurate measurements of both the induced voltage $U$ and the voltage drop across the standard resistor $R$. In conjunction with this, we have established high-precision digital voltage meters (DVMs) for the meticulous measurement of residual voltage.

    \item [-] \textit{Resistance standard, $R$.}
    In addition to the voltage reference, the Kibble balance necessitates a resistance standard, denoted as $R$, for precise current measurements, featuring typical uncertainties at the $10^{-9}$ level. In the context of the Tsinghua Kibble balance, a highly stable $100\,\Omega$ standard resistor sourced from Alpha Electronics (HRU-101) fulfills this role as the designated resistance standard. The standard resistor is meticulously housed within an air bath with temperature stability of approximately 1\,mK. Recently, the initial calibration of this resistor against the quantum Hall resistance has been conducted, yielding a calibration uncertainty of $5\times10^{-9}$.

    \item [-] \textit{Frequency reference, $f$.}
    Both the Josephson system and the optical measurement system necessitate a high-precision frequency reference. To fulfill this requirement, a frequency standard (specifically, the FS740 with a Rubidium timebase) has been integrated into the laboratory setup. Leveraging the GPS for synchronization, the frequency standard attains a remarkable long-term accuracy on the order of $10^{-13}$.
    
\end{itemize}

\section{Summary and outlook}
\label{sec:04}

In conclusion, the Tsinghua University tabletop Kibble balance project aims to develop a compact, highly accurate, and cost-effective mass calibration instrument, focusing on enhancing measurement precision while maintaining a small form factor. Significant progress has been made in the design and implementation of key components, including a novel magnet system with enhanced field uniformity, a custom-designed weighing unit with superior position measurement sensitivity, and a stable, dual-stage current source. These advancements collectively contribute to achieving the project's goal of accurately calibrating masses from 10\,g to 1\,kg with a measurement uncertainty below 50\,$\upmu$g, thereby making the Kibble balance more accessible and practical for widespread use in metrology. The subsequent phase involves the integration of these components or subsystems into a cohesive whole. The anticipation is the creation of a precision-oriented and robust mass-realization instrument in the near future.

\section*{Acknowledgment}
The authors would like to acknowledge the BIPM Kibble balance group and Dr. Tao Zeng from China Jiliang University for valuable discussions on the optical system design. Shisong Li would like to thank Dr. Lushai Qian from China Jiliang University and Dr. Yang Dong from Shenyang Chuangyuan Measuring Instrument Co., Ltd. for discussions on the weighing unit design.

\end{document}